\documentclass[12pt,preprint]{aastex}

\usepackage{amsmath}
\usepackage{natbib}
\usepackage{graphicx}
\usepackage{color}
\usepackage[breaklinks,colorlinks,citecolor=blue]{hyperref}
\usepackage{epsfig}
\usepackage{epstopdf}

\usepackage{mathrsfs}

\begin{document}
\title{Perfect fluid dark matter influence on thermodynamics and phase transition for a Reissner-Nordstrom-anti-de Sitter black hole}

\author{
  Zhaoyi Xu\altaffilmark{1,2,3,4},
  Xian Hou\altaffilmark{1,3,4},
  Jiancheng Wang\altaffilmark{1,2,3,4},
  and Yi Liao\altaffilmark{5,6}
 }

\altaffiltext{1}{Yunnan Observatories, Chinese Academy of Sciences, 396 Yangfangwang, Guandu District, Kunming, 650216, P. R. China; {\tt zyxu88@ynao.ac.cn,xhou@ynao.ac.cn,jcwang@ynao.ac.cn}}
\altaffiltext{2}{University of Chinese Academy of Sciences, Beijing, 100049, P. R. China}
\altaffiltext{3}{Key Laboratory for the Structure and Evolution of Celestial Objects, Chinese Academy of Sciences, 396 Yangfangwang, Guandu District, Kunming, 650216, P. R. China}
\altaffiltext{4}{Center for Astronomical Mega-Science, Chinese Academy of Sciences, 20A Datun Road, Chaoyang District, Beijing, 100012, P. R. China}
\altaffiltext{5}{Department of Physics, National University of Defense Technology, Changsha, 410073, P. R. China; {\tt liaoyitianyi@gmail.com}}
\altaffiltext{6}{Interdisciplinary Center for Quantum Information, National University of Defense Technology, Changsha, 410073, P. R. China}

\shorttitle{Perfect fluid dark matter influence on thermodynamics and phase transition for a Reissner-Nordstrom-anti-de Sitter black hole}
\shortauthors{Z Y. Xu et al.}

\begin{abstract}
Based on Reissner-Nordstrom-anti-de Sitter(RN-AdS) black hole surrounded by perfect fluid dark matter, we study the thermodynamics and phase transition by extending the phase space defined by the charge square $Q^{2}$ and the conjugate quantity $\psi$, where $\psi$ is a function of horizon radius. The first law of thermodynamics and the equation of state are derived in the form $Q^{2}=Q^{2}(T,\psi)$. By investigating the critical behaviour of perfect fluid dark matter around Reissner-Nordstrom-anti-de Sitter black hole, we find that these thermodynamics system are similar to Van der Waals system, and can be explained by mean field theory. We also explore the Ruppeiner thermodynamic geometry feature and their connection with microscopic structure. We find that in extended phase space there existence singularity points of Ruppeiner curvature and it's could  explained as phase transitions.
\end{abstract}

\keywords {Perfect fluid dark matter, RN-AdS black hole, Thermodynamics and phase transition, Critical phenomenon, Ruppeiner thermodynamic geometry}

\section{INTRODUCTION}
Black hole thermodynamics are one of the most important topics in modern physics research and have been widely studied in recent years. The laws
of black hole dynamics and thermodynamics were analyzed by Bekenstein and Hawking (\cite{1972NCimL...4..737B,1974PhRvD...9.3292B,1973PhRvD...7.2333B,1971PhRvL..26.1344H,1974Natur.248...30H}). The four laws of black hole thermodynamics have been discussed (\cite{1973CMaPh..31..161B}). Since the Hawking-Page phase transition was discovered, phase transitions have become a important topic in the black hole area. There are lots of  work on the phase transition of different black holes, such as Reissner-Nordstrom black hole, Kerr black hole and Kerr-Newman black hole (\cite{2000CQGra..17..399C,1989CQGra...6.1909D,1978RPPh...41.1313D,1977RSPSA.353..499D}). These work have also been generalized to other black holes or applied to general situations (\cite{1982CMaPh..87..577H,1977MNRAS.180..379H,2016PhLB..760.112, 2017ApJ..835.247,2016arXiv160804176M,2016arXiv160906224R}). Recently, several studies have considered the cosmological constant as a dynamical variable which is similar to thermodynamical pressure (\cite{2012JHEP...07..033K}). Utilizing this method, some work have obtained the phase transition in AdS-black hole, in which the analogy between the critical behaviours of the Van der Waals gas and the RN-AdS black hole have been found (\cite{2016EPJC...76...73B,2015EPJC...75...71B,2012ChPhL..29j0401B,2013ChPhL..30i0402B,2016EPJC...76..304C}).

From recent observations, we know that our universe is dominated by Dark Energy and Dark Matter (\cite{2011ApJS..192...18K,2004ApJ...607..665R}). The dark energy makes the universe to be in accelerated expansion, and its state of equation is very close to the cosmological constant or the  vacuum energy (\cite{2003RvMP...75..559P}). But the dynamics of dark energy zre more like quintessence or other dynamical dark energy in behaviour (\cite{2006IJMPD..15.1753C}). The dark energy with quintessence could affect the black hole spacetime (\cite{2005MPLA...20..561S}).
For the schwarzschild black hole in quintessence field, the modified black hole metric has been obtained by \cite{2003CQGra..20.1187K}. Recently the rotational quintessence black hole and Kerr-Newman-AdS black hole solutions have also been obtained (\cite{2015arXiv151201498T,2017PhRvD...95..06415S}). For the schwarzschild black hole surrounded by quintessence matter, the thermodynamics and phase transition have been discussed in \cite{2014MPLA...2950057T,2016Ap&SS.361..161G,2016NuPhB.903...10G}. For rotational black hole surrounded by quintessence case, the thermodynamics and phase transition have been studied recently by  \cite{2016arXiv161000376X}. The Rerssner-Nordstrom black hole and Rerssner-Nordstrom-dS black hole have been studied by \cite{2016NuPhB.903...10G,2011ChPhL..28j0403W,2013ChPhB..22c0402W,2012GReGr..44.2181T,2015Ap&SS.357....8M,2016arXiv160606070M}, the thermodynamics (\cite{2016arXiv160307748P,2017arXiv170407720C}) and phase transition through holography framework for Reissner-Nordstrom-AdS black hole have been investigated (\cite{2015arXiv151208855Z}). Following these works, the cold dark matter around black hole in phantom field background have been obtained by \citep[][and references therein]{2012PhRvD..86l3015L}.  In this paper, we study the thermodynamics and phase transition of Reissner-Nordstrom-AdS black hole surround by perfect fluid  dark matter.

This paper focuses on the study of the influence of perfect fluid dark matters on thermodynamics and phase transition for Reissner-Nordstrom-anti-de Sitter black hole. In section 2, we introduce the Reissner-Nordstrom-anti-de Sitter black hole with perfect fluid dark matter background. In section 3, we study the thermodynamical features, the equation of state in $(Q^{2},\psi)$ space and the critical behaviour. In section 4, we study the Ruppeiner thermodynamic geometry and their connection with microscopic structure for these black hole systems. We summarize our results in Section 5.

\section{THE SPACETIME OF PERFECT FLUID DARK MATTER AROUND REISSNER-NORDSTROM-ADS BLACK HOLE}
We consider the dark matter field minimally coupled to gravity, electromagnetic field and cosmological constant (\cite{2012PhRvD..86l3015L,2003gr.qc.....3031K,2005CQGra..22..541K})
\begin{equation}
S=\int d^{4}x\sqrt{-g}(\dfrac{1}{16\pi G}R-\dfrac{1}{8\pi G}\Lambda+\dfrac{1}{4}F^{\mu\nu}F_{\mu\nu}+\mathcal{L}_{DM}),
\label{Action1}
\end{equation}
where $G$ is the Newton gravity constant, $\Lambda$ is the cosmological constant, $F_{\mu\nu}$ is the Faraday tensor of electromagnetic field,  $\mathcal{L}_{DM}$ is the dark matter Lagrangian density, this dark matter can be any dark matter model. By variational we obtain the field equation from action principle as
\begin{equation}
R_{\mu\nu}-\dfrac{1}{2}g_{\mu\nu}R+\Lambda g_{\mu\nu}=-8\pi G (\bar{T_{\mu\nu}}+T_{\mu\nu}(DM))=-8\pi G T_{\mu\nu},$$$$
F^{\mu\nu}_{~~~;\nu}=0,~~~ F^{\mu\nu;\alpha}+F^{\nu\alpha;\mu}+F^{\alpha\mu;\nu}=0.
\label{einstein equation 1}
\end{equation}
Where $\bar{T_{\mu\nu}}$ is the energy-momentum tensors of ordinary matter and $T_{\mu\nu}(DM)$ is the energy-momentum tensors of dark matter. In the case of black holes surrounded by dark matter, we assume that dark matter is perfect fluid. The energy-momentum tensors can be written as $T^{t}_{~~t}=-\rho, T^{r}_{~~r}=T^{\theta}_{~~\theta}=T^{\phi}_{~~\phi}=p$ (where $T^{\mu}_{~~\mu}=g^{\mu\nu}T_{\mu\nu}$). In addition, for the simpliest case, we assume $T^{r}_{~~r}=T^{\theta}_{~~\theta}=T^{\phi}_{~~\phi}=T^{t}_{~~t}(1-\delta)$, where $\delta$ is a constant. We refer to such dark matter as the "perfect fluid dark matter" in this work.

For the Reissner-Nordstrom-AdS spacetime metric in perfect fluid DM, the black hole solution is \citep{2003gr.qc.....3031K,2012PhRvD..86l3015L}
\begin{equation}
ds^{2}=-f(r)dt^{2}+f^{-1}(r)dr^{2}+r^{2}(d\theta^{2}+sin^{2}\theta d\phi^{2}),
\label{solution1}
\end{equation}
where
\begin{equation}
f(r)=1-\dfrac{2M}{r}+\dfrac{Q^{2}}{r^{2}}+\dfrac{1}{3}\Lambda r^{2}+\dfrac{\alpha}{r}ln(\dfrac{r}{\mid\alpha \mid}),
\label{solution2}
\end{equation}
where $\alpha$ is a parameter describing the intensity of the perfect fluid DM, $M$ is the black hole mass and $Q$ is the charge of black hole.
This solution corresponds to a specific case of the general solution in \cite{2003gr.qc.....3031K} and \cite{2012PhRvD..86l3015L}.
It is interesting to note that this black hole solution implies that the rotational velocity is asymptotically flat in the equatorial plane, which could explain the observed rotation curves in spiral galaxies \citep{2003gr.qc.....3031K,2012PhRvD..86l3015L}.

\section{THERMODYNAMICS OF DARK MATTER AROUND REISSNER-NORDSTROM-ADS BLACK HOLE}
In section 2, we have obtained the metric of spherically symmetric Reissner-Nordstom-AdS black hole in perfect fluid dark matter. These black holes have three horizons which are Cauchy horizon $r_{-}$, event horizon $r_{+}$ and cosmological horizon $r_{\Lambda}$. In this work, we always use event horizon $r_{+}$. The black hole mass $M$ can be expressed in event horizon $r_{+}$ as
\begin{equation}
M=\dfrac{r_{+}}{2}+\dfrac{Q^{2}}{2r_{+}}+\dfrac{1}{6}\Lambda r_{+}^{3}+\dfrac{\alpha}{2}ln(\dfrac{r_{+}}{\mid\alpha \mid}).
\label{Mass1}
\end{equation}
The semi-hawking temperature and the entropy are given by
\begin{equation}
T=\dfrac{1}{4\pi}\dfrac{df(r)}{dr}\mid_{r=r_{+}}=\dfrac{r_{+}(2M+\alpha(1-ln(\dfrac{r_{+}}{\mid\alpha \mid}))+\dfrac{2}{3}\Lambda r^{3}_{+})-2Q^{2}}{4\pi r^{3}_{+}}
=\dfrac{r_{+}(\alpha+r_{+}+\Lambda r^{3}_{+})-Q^{2}}{4\pi r^{3}_{+}},
\label{temperature1}
\end{equation}

\begin{equation}
S=\int^{r_{+}}_{0}\dfrac{1}{T}(\dfrac{\partial M}{\partial r_{+}})dr_{+}=\pi r^{2}_{+}.
\label{entropy1}
\end{equation}

Now we study the thermodynamical properties of the black hole with perfect fluid dark matter by extending to new phase space. This phase space is constructed by the entropy $S$, the perfect fluid dark matter density $\alpha$, the charge square $Q^{2}$ and the pressure $P=-\dfrac{\Lambda}{8\pi}$ corresponding
to the cosmological constant $\Lambda$. Therefore the black hole mass can be expressed as
\begin{equation}
M(S,Q^{2},P,\alpha)=\dfrac{1}{2}\sqrt{\dfrac{S}{\pi}}+\dfrac{Q^{2}}{2}\sqrt{\dfrac{\pi}{S}}-\dfrac{4}{3}PS\sqrt{\dfrac{S}{\pi}}+\dfrac{\alpha}{2}ln(\dfrac{1}{\mid\alpha \mid}\sqrt{\dfrac{S}{\pi}}).
\label{Mass2}
\end{equation}
The intensive parameters are defined by
\begin{equation}
T=\dfrac{\partial M}{\partial S}\mid_{P,Q^{2},\alpha},~~~~ \psi=\dfrac{\partial M}{\partial Q^{2}}\mid_{S,P,\alpha},$$$$
V=\dfrac{\partial M}{\partial P}\mid_{S,Q^{2},\alpha},~~~~\Pi=\dfrac{\partial M}{\partial \alpha}\mid_{S,Q^{2},P},
\label{Thermodynamics1}
\end{equation}
where $T$ denotes the temperature, the new physical quantity $\psi$ is  related to the specific volume as $\psi=1/v$, where $v=2r_{+}$. In thermodynamical space, the volume is $V=4\pi r^{3}_{+}/3$ and the quantity $\Pi=\dfrac{1}{2}ln(\dfrac{r_{+}}{\mid\alpha \mid})$. The generalized first law of black hole thermodynamics in this extended phase space is expressed by
\begin{equation}
dM=TdS+\psi dQ^{2}+VdP+\Pi d\alpha,
\label{Thermodynamics2}
\end{equation}
and the generalized Smarr formula is given by
\begin{equation}
M=2TS+\psi Q^{2}-2VP+\Pi\alpha.
\label{Thermodynamics3}
\end{equation}
For the first law of black hole thermodynamics, the term $\psi dQ$ becomes $\psi dQ^{2}$ in this phase space, where $\psi$ represents the electric potential. This change leads to the interesting behaviour with $dM=SdT+\psi dQ$ in formal phase space. When the perfect fluid dark matter is around black hole, the phase transition of black hole occurs in $(P,v)$ plane. We can discuss this phase transition in $(Q^{2},\psi)$ plane, including the critical point, Gibbs free energy and critical exponents under the effect of  perfect fluid dark matter.
Through calculations, we obtain the state equation $Q^{2}(T,\psi)$ as
\begin{equation}
Q^{2}=r_{+}(\alpha+r_{+}+\Lambda r^{3}_{+}-4\pi r^{2}_{+}T)
=\dfrac{1}{2\psi}[\alpha+\dfrac{1}{2\psi}+\dfrac{3}{8\psi^{3}l^{2}}-\dfrac{\pi}{\psi^{2}}T],
\label{Thermodynamics4}
\end{equation}
where $l^{2}=3/\Lambda$. The above equation describes the behaviours of $Q^{2}$ for different $T$,$\psi$ and $\alpha$.

$Q^{2}$ and $\psi$ also satisfy the Maxwell equal area theorem given by \cite{2013arXiv1310.2186S}
\begin{equation}
\oint\psi dQ^{2}=0.
\label{Thermodynamics4}
\end{equation}
For $T>T_{c}$, where $T_{c}$ critical temperature, there is an inflection point which is similar to the Van der Waals system. From general method, the coordinates of the critical point is determined by two equations
\begin{equation}
\dfrac{\partial^{2}Q^{2}}{\partial\psi^{2}}\mid_{T_{c}}=0,~~~~\dfrac{\partial Q^{2}}{\partial\psi}\mid_{T_{c}}=0.
\label{Thermodynamics5}
\end{equation}
We then obtain the following equations
\begin{equation}
\dfrac{15}{4\psi^{3}_{c}l^{2}}+\alpha+\dfrac{3}{2\psi_{c}}-\dfrac{6\pi}{\psi^{2}_{c}}T_{c}=0,$$$$
-\dfrac{3}{2\psi^{3}_{c}l^{2}}-\alpha-\dfrac{1}{\psi_{c}}+\dfrac{3\pi}{\psi^{2}_{c}}T_{c}=0.
\label{Thermodynamics61}
\end{equation}
We then get the equation of critical $\psi_{c}$ and $T_{c}$ as
\begin{equation}
\alpha\psi^{3}_{c}+\dfrac{1}{2}\psi^{2}_{c}-\dfrac{3}{4 l^{2}}=0,$$$$
T_{c}=\dfrac{\psi^{2}_{c}}{3\pi}(\dfrac{3}{2\psi^{3}_{c}l^{2}}+\alpha+\dfrac{1}{\psi_{c}}).
\end{equation}
The universal number as
\begin{equation}
\rho_{c}=Q^{2}_{c}T_{c}\psi_{c}=\dfrac{\psi^{2}_{c}}{3\pi}(\alpha+\dfrac{1}{4}+\dfrac{3}{2\psi^{3}_{c}l^{2}})(\dfrac{\alpha}{3}+\dfrac{1}{12\psi_{c}}-\dfrac{1}{16\psi^{3}_{c}l^{2}}),
\label{Thermodynamics7}
\end{equation}
where $0<\alpha<2$ for perfect fluid dark matter. When $\alpha=0$, we have the following values
\begin{equation}
T_{c}=\dfrac{\sqrt{6}}{3\pi l},~~~~ Q^{2}_{c}=\dfrac{l^{2}}{36},~~~~ \psi_{c}=\sqrt{\dfrac{3}{2l^{2}}},
\label{Thermodynamics6}
\end{equation}
which have been obtained by \cite{2017PhLB..768..235D}.

Now we study the critical behaviour near the phase transition with perfect fluid dark matter in new phase space. We first define
$\psi_{r}=\psi/\psi_{c}$, $Q^{2}_{r}=Q^{2}/Q^{2}_{c}$ and $T_{r}=T/T_{c}$, we then get $\psi_{r}=1+\chi$, $Q^{2}_{r}=1+\varrho$ and $T_{r}=1+t$,
where $\chi,\varrho$ and $t$ represent the deviations away from critical points.
The thermodynamics quantities are defined as
\begin{equation}
C_{\psi}=\mid t\mid^{-a},$$$$
\eta=\mid t\mid^{b},$$$$
\kappa_{T}=\mid t\mid^{-c}$$$$
\mid Q^{2}-Q^{2}_{c}\mid=\mid \psi-\psi_{c}\mid^{d}
\label{Thermodynamics9}
\end{equation}
where $a, b, c$ and $d$ are the critical exponents.
The critical exponent $a$ is derived by fixing potential $\psi$ for heat capacity as
\begin{equation}
C_{\psi}=T\dfrac{\partial S}{\partial T}\mid_{\psi}=0, ~~~a=0.
\label{Thermodynamics10}
\end{equation}
The critical exponent $b$ is derived from $Q^{2}$ as
\begin{equation}
Q^{2}_{r}=\dfrac{1}{2Q^{2}_{c}\psi_{c}\psi_{r}}(\alpha+\dfrac{1}{2\psi_{r}\psi_{c}}+\dfrac{3}{8l^{2}\psi^{3}_{r}\psi^{3}_{c}}-\dfrac{\pi T_{c}}{\psi^{2}_{r}\psi^{2}_{c}}T_{r})
=\dfrac{F_{1}(\psi_{c})}{\psi_{r}}+\dfrac{F_{2}(\psi_{c})}{\psi^{2}_{r}}+\dfrac{F_{3}(\psi_{c})}{\psi^{3}_{r}}+\dfrac{F_{4}(\psi_{c})}{\psi^{4}_{r}}T_{r},
\label{Thermodynamics11}
\end{equation}
where $F_{1}(\psi_{c})=\alpha/(2Q^{2}_{c}\psi_{c}), F_{2}(\psi_{c})=1/(4Q^{2}_{c}\psi^{2}_{c}), F_{3}(\psi_{c})=3/(16Q^{2}_{c}\psi^{4}_{c}l^{2}), F_{4}(\psi_{c})=-\pi T_{c}/(2Q^{2}_{c}\psi^{3}_{c})$, and the critical points $\psi_{c},Q^{2}_{c},T_{c}$ are functions of $\alpha, l^{2}$.
In order to obtain critical exponent, we expand the equation near the critical point using $\psi_{r}=1+\chi$, $T_{r}=1+t$ and $Q^{2}_{r}=1+\varrho$, and obtain
\begin{equation}
\varrho=F_{4}(\psi_{c})t-4F_{4}(\psi_{c})t\chi-(F_{1}(\psi_{c})+4F_{2}(\psi_{c})+4F_{3}(\psi_{c})+50F_{4}(\psi_{c}))\chi^{3}+ high~order~term.
\label{Thermodynamics12}
\end{equation}
Through differentiating Eq.\ref{Thermodynamics12} with respect to $\chi$ and $\chi$, and use equation (\ref{Thermodynamics4}), we obtain $\chi_{s}=-\chi_{l}=\sqrt{-4F_{4}(\psi_{c})t/(F_{1}(\psi_{c})+4F_{2}(\psi_{c})+4F_{3}(\psi_{c})+50F_{4}(\psi_{c}))}$, where $l$ and $s$ represent large and small black hole phase respectively. We then find that
\begin{equation}
\mid\chi_{s}-\chi_{l}\mid=2\chi_{s}=\sqrt{\dfrac{-16F_{4}(\psi_{c})}{F_{1}(\psi_{c})+4F_{2}(\psi_{c})+4F_{3}(\psi_{c})+50F_{4}(\psi_{c})}}t^{\dfrac{1}{2}},~~~b=\dfrac{1}{2}.
\label{Thermodynamics13}
\end{equation}
The critical exponent $c$ is derived from isothermal compressibility coefficient $\kappa_{T}$ as
\begin{equation}
\kappa_{T}=\dfrac{\partial \psi}{\partial Q^{2}}\mid_{T}\propto\dfrac{\psi_{c}}{-4F_{4}(\psi_{c})Q^{2}_{c}t},~~~c=1,
\label{Thermodynamics14}
\end{equation}
The critical exponent $d$ is obtained from equation (\ref{Thermodynamics12}) as
\begin{equation}
\varrho\mid_{t=0}=-(F_{1}(\psi_{c})+4F_{2}(\psi_{c})+4F_{3}(\psi_{c})+50F_{4}(\psi_{c}))\chi^{3},~~~d=3.
\label{Thermodynamics15}
\end{equation}

From the above analysis, we find that these critical exponents resemble those in Van der Waals system, implying that
the critical phenomenon can be explained by mean field theory (\cite{2012JHEP...07..033K}).
These critical exponents satisfy the scale symmetry given by
\begin{equation}
a+2b+c=2,$$$$
a+b(d+1)=2,$$$$
c(d+1)=(2-a)(d-1),$$$$
c=b(d-1).
\label{SM1}
\end{equation}

The perfect fluid dark matters around black holes could explain
the formation of supermassive black holes in approximate stationary situation. In the past decades, some high redshift quasars have been discovered and the centre supermassive black hole with the mass beyond 10 billion of solar masses. Usually such black hole is difficult to form in the Universe less than one billion years old. Because perfect fluid dark matter black holes satisfy the first law of thermodynamics, the perfect fluid dark matter could accelerate the formation of supermassive black holes.

\section{GEOTHERMODYNAMICS OF DARK MATTER AROUND REISSNER-NORDSTROM-ADS BLACK HOLE}
In the black hole thermodynamics, the thermodynamic geometry method is a usual tool to understand the property of black hole thermodynamics system.
In the work, we use the Ruppeiner metric to study the thermodynamical effect of the perfect fluid dark matter on the
microscopical structure of Reissner-Nordstrom-AdS black hole (\cite{2010JHEP...07..082S}). We define the metric on $(M,Q^{2})$ space given by
\begin{equation}
g_{\mu\nu}(R)=\dfrac{1}{T}\dfrac{\partial^{2}M}{\partial X^{\mu}\partial X^{\nu}}={\left(\begin{array}{cc}
\dfrac{1}{T}\dfrac{\partial^{2} M}{\partial S^{2}}&\dfrac{1}{T}\dfrac{\partial^{2} M}{\partial S \partial Q^{2}}\\
\dfrac{1}{T}\dfrac{\partial^{2} M}{\partial S \partial Q^{2}}&\dfrac{1}{T}\dfrac{\partial^{2} M}{\partial Q^{4}}
\end{array}
\right)}
\label{TG1}
\end{equation}
where $X^{\mu}=(S,Q^{2})$.

From Equations (\ref{Mass1}) and (\ref{temperature1}), we find that $M(S,Q^{2})$ and $T(S,Q^{2})$ in $(M,Q^{2})$ space are given by
\begin{equation}
M(S,Q^{2})=\dfrac{1}{2}\sqrt{\dfrac{S}{\pi}}+\dfrac{Q^{2}}{2}\sqrt{\dfrac{\pi}{S}}+\dfrac{1}{2l^{2}\pi}S\sqrt{\dfrac{S}{\pi}}+\dfrac{\alpha}{2}ln(\dfrac{1}{\mid\alpha \mid}\sqrt{\dfrac{S}{\pi}}),
\label{TG2}
\end{equation}

\begin{equation}
T(S,Q^{2})=\dfrac{\alpha}{4S}+\dfrac{3}{4\pi l^{2}}\sqrt{\dfrac{S}{\pi}}+\dfrac{1}{4\pi}\sqrt{\dfrac{\pi}{S}}-\dfrac{Q^{2}}{4S}\sqrt{\dfrac{\pi}{S}}.
\label{TG3}
\end{equation}

From geometry, the Ricci scalar can be calculated for perfect fluid dark matter around Reissner-Nordstrom-AdS black holes by
\begin{equation}
R(RN-AdSDM)=R^{\mu\nu\rho\sigma}R_{\mu\nu\rho\sigma}=\dfrac{H^{2}}{(-l^{2}\pi^{2}Q^{2}-2\alpha l^{2}\sqrt{\pi^{3} S}+l^{2}\pi S+3S^{2})^{2}}$$$$
=\dfrac{H^{2}}{(-l^{2}\pi^{2}Q^{2}-\dfrac{\alpha l^{2}\pi^{2}}{\psi}+\dfrac{l^{2}\pi^{3}}{4\psi^{2}}+\dfrac{3\pi^{2}}{16\psi^{4}})^{2}},
\label{TG4}
\end{equation}
where $H$ are function of $\alpha,l^{2},Q^{2}$ and $S$. From above equation, we find that the phase transition occurs when the following condition is satisfied
\begin{equation}
-l^{2}\pi^{2}Q^{2}-\dfrac{\alpha l^{2}\pi^{2}}{\psi}+\dfrac{l^{2}\pi^{3}}{4\psi^{2}}+\dfrac{3\pi^{2}}{16\psi^{4}}=0.
\label{TG5}
\end{equation}

We know that the sign of the Ricci scalar $R$ can be explained by intermolecular interaction in thermodynamical system.
The positive sign refers to the repulsive interaction between the constituents of the thermodynamical system, while
the negative sign refers to the attractive interaction between the constituents of the thermodynamical system \citep[][and references therein]{2017arXiv170407720C}.
For perfect fluid dark matter around Reissner-Nordstrom-AdS black hole, the interactions are absent in the thermodynamical
system for a null Ricci scalar $R$, which is similar to that in classical ideal gas (\cite{1979PhRvA..20.1608R}).

\section{SUMMARY}
In the paper, we study the perfect fluid dark matter influence on thermodynamics and phase transition of Reissner-Nordstrom-AdS black hole by extending phase space defined by the charge square $Q^{2}$ and conjugate quantity $\psi$. The first law of thermodynamics and the equation of state are derived in the form of $Q^{2}=Q^{2}(T,\psi)$. We analyze the critical behaviour of dark matter around Reissner-Nordstrom-AdS black hole and find that these thermodynamics system resemble the Van der Waals system which can be explained by mean field theory. We also find that the critical exponents satisfy
the scale law of thermodynamical system. Using Ruppeiner thermodynamic geometry, we study the geometric property of perfect fluid dark matter around black holes.
We find that in extended phase space, some singular points appear on the Ruppeiner curvature, which can be explained as the critical points of
phase transitions.

The Reissner-Nordstrom-AdS black hole surrounded by dark matters could appear in the Universe. In the future work we plan to study the observed
effects of perfect fluid dark matter on black holes and the influence of perfect fluid dark matter on gravitational
lensing and the evolution of dark matter in the universe.

\acknowledgments
We acknowledge the anonymous referee for a constructive report that significantly improved this paper. We acknowledge the financial support
from the National Natural Science Foundation of China 11573060 and  11661161010.

\end{document}